\documentclass[seceq]{ptptex}

\usepackage{graphicx}

\title{
Light scalar mesons as tetraquarks within QCD Sum Rules%
}

\author{
Hee-Jung \textsc{Lee}\footnote{ e-mail address:
hjl@chungbuk.ac.kr}
}

\inst{
Department of Science Education, Chungbuk National University,
Cheongju Chungbuk 361-763, South Korea\\
}

\abst{
We examine the interpretation of the light scalar meson nonet as tetraquark
states using QCD sum rules. With the interpolating current for the tetraquark
states composed of scalar diquark and scalar antidiquark, first, we construct
the QCD sum rules by means of the operator product expansion up to the operators
of dimension 8 and show that there is no evidence of the coupling of the tetraquark
states to the light scalar meson nonet. In order to have a stable sum rule,
we propose a ``good" interpolating current for the tetraquarks based on
chirality arguments which includes scalar and pseudoscalar diquark--antidiquarks
with equal weights. In particular, for the lowest tetraquark $\sigma$--meson,
we perform detail analysis of the QCD sum rule and obtain mass for the $\sigma(600)$
around 780 MeV.
}

\begin{document}

\maketitle


\section{Introduction}

It is well known that some properties of the light scalar meson nonet
stimulated to interpret them as various tetraquark states~\cite{Amsler04}.
It is important to get a justification for such tetraquark states
from QCD. First we study the light scalar meson nonet with quark
content of the scalar diquark and the scalar antidiquark within
the QCD sum rules. Since the correlator for the tetraquarks for
the sum rules has higher energy dimension than that for ordinary baryons,
the operator product expansion (OPE) must be considered up to higher operators
than those for ordinary baryons. We include the operators in OPE up to
dimension 8 with the factorization hypothesis. It is shown that the contributions
from the dimension 8 condensates are unexpectively large to become
dominant in the sum rules. Moreover their negative contributions to
the sum rule break down the physical meaning of the sum rule.

In order to find a good interpolating current for which the large
contributions from the high dimension operators could be canceled out
in the sum rules, we study the current based on the instanton model for
the QCD vacuum for two flavors. Since the $\sigma(600)$ has vacuum
quantum numbers, this current is expected to have a strong coupling to $\sigma$ state.
An additional advantage of using instanton originated current lies in the fact that
there is cancelation of the high dimension operator contributions in
the OPE in the self dual fields~\cite{smilga}.
With this inspection, we propose a ``good" interpolating current for $\sigma(600)$.
We show that this current has very peculiar chirality structure and leads to the
cancelation of large high dimension operator contributions and some dangerous
instanton contributions in the sum rules. With the OPE up to operators of
dimension 10, we fit the mass of $\sigma(600)$ by applying the two resonance
approximation to the phenomenological side of the sum rule in order to avoid
the well--known problem of strong dependence of results on the value of the threshold
for multiquark systems~\cite{thres}.

\section{QCD sum rules for the light scalar mesons with scalar diquarks}
In the picture of the tetraquark states, the light scalar meson nonet is
generated by the diquark in the $\overline{\mathbf 3}_f$ and the antidiquark
in the ${\mathbf 3}_f$, where the subscript $f$ stands for flavor.
Accepting the argument from the constituent quark model that two quarks in the
scalar channel feel the strongest attraction by the perturbative one--gluon
exchange~\cite{jaffe2} and the non--perturbative instanton dynamics~\cite{shuryak},
the scalar diquark (antidiquark) should belong to
$\overline{\mathbf 3}_c$ (${\mathbf 3}_c)$ in color space and to spin--zero state
by Fermi statistic. From this structure, the interpolating current for
the scalar nonet can be written as
\begin{equation}
J_S=N_S\epsilon_{abc}\epsilon_{ade}(q_{1b}^{T}\Gamma q_{2c})
(\bar{q}_{3d}\overline{\Gamma} \bar{q}_{4e}^T)\ ,
\end{equation}
where $\Gamma=C\gamma_5$ and $\overline{\Gamma}=\gamma^0\Gamma^\dagger\gamma^0$.
Here $N_S$ is the normalization constant, the indices $a, b, c,\cdots$
denote color, and the subscripts $1,2,3,4$ are introduced for
flavor (See Ref.~\cite{Lee05} for more details).

\begin{figure}[tbh]
\centering
\includegraphics[width=5.0cm]{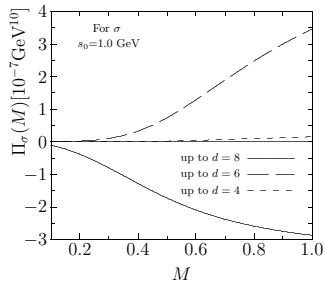}
\caption{LHS of the QCD sum rule for the $\sigma(600)$
using the scalar diquark--antidiquark interpolating current.}
\label{Pisig}
\end{figure}

Within the narrow one resonance approximation in the phenomenological part
of the sum rule, the Borel transform yields the following QCD sum rules up to
the operators of energy dimension 8 for the scalar meson nonet :
\begin{eqnarray}
&&C_{0}^{S}{\cal O}_0M^{10}E_4(M)
+C_{4,i}^S{\cal O}_{4,i}M^{6}E_{2,i}(M)
+C_{6,i}^S{\cal O}_{6,i}M^{4}E_{1,i}(M)
\nonumber\\
&&+C_{8,i}^S{\cal O}_{8,i}M^{2}E_{0,i}(M)
=2f_{S}^2m_{S}^8e^{-m_{S}^2/M^2}\ ,
\label{sumrule}
\end{eqnarray}
where $M$ is the Borel mass. The decay constant and the mass of
the mesons of the scalar nonet are defined by
$\langle0|J_S|S\rangle=\sqrt{2}f_Sm_S^4\ .$
The first index in the coefficients $C_{d,i}$ denotes the dimension in power
of energy of the associated operators ${\cal O}_{d,i}$. Explicit forms of
the sum rules are given in Ref.~\cite{Lee05}. As shown in Fig.~\ref{Pisig}
for $\sigma(600)$, the negative contributions from the operators
$\langle\bar{q}q\rangle\langle\bar{q}ig\sigma\cdot Gq\rangle$
where $q=u,d,s$ in the dimension 8 operators dominate in the sum rule and make
the left hand side (LHS) of the sum rule to become negative definite.
The same situation happens in the sum rules for other members in the nonet.
Moreover if we include the contributions from the direct instantons,
we have a worse situation. This means that the QCD sum rules with the tetraquark
interpolating currents consisting of the scalar diquark and the scalar antidiquark
cannot be used to derive the properties of the light scalar meson nonet.

\section{QCD sum rule for $\sigma(600)$ with a current
based on the instanton picture for the QCD vacuum}
Generally, there are two possibilities in getting the color singlet from
a system consisting of the diquark and the antidiquark. One is that the diquark
belongs to $\overline{\mathbf 3}_c$ and the antidiquark to ${\mathbf 3}_c$.
Another is that the diquark belongs to ${\mathbf 6}_c$ and the antidiquark
to $\overline{\mathbf 6}_c$. These two possibilities inform us that there
are five types of diquarks : scalar($S$),
pseudoscalar($P$), vector($V$), axial vector($A$), and tensor($T$).
Therefore, in general, the interpolating current for the light scalar mesons
may have the form
\begin{equation}
J_S=\alpha J_S^{S\bar{S}}+\beta J_S^{P\bar{P}}
+\gamma J_S^{T\bar{T}}+\omega J_S^{V\bar{V}}+\eta J_S^{A\bar{A}}\ ,
\end{equation}
where $J_S^{i\bar{i}}$ means the current consisting of the appropriate type of
diquark ($i$) and antidiquark ($\bar{i}$). To fix the coefficients, we will use
the instanton induced diquark antidiqaurk vertices which can be obtained from
the famous t'Hooft instanton induced quarks interaction for the two flavor
case~\cite{Rapp98} :
\begin{eqnarray}
{\cal L}=\frac{G}{4(N_c^2-1)}\bigg[
\frac{2N_c-1}{2N_c}\bigg((\bar{\psi}\tau^-_\mu\psi)^2
+(\bar{\psi}\gamma_5\tau^-_\mu\psi)^2\bigg)
+\frac{1}{4N_c}(\bar{\psi}\sigma_{\rho\sigma}\tau^-_\mu\psi)^2\bigg]\ .
\label{lag}
\end{eqnarray}
By using the Fierz transform, we can rewrite the Lagrangian in terms of
the diquarks
\begin{eqnarray}
{\cal L}
&=&\frac{G}{2N_c(N_c-1)}\ \epsilon_{abc}\epsilon_{ade}\bigg(
(u^T_b\Gamma_{S}d_{c})(\bar{u}_d\overline\Gamma_{S}\bar{d}^T_{e})
-(u^T_b\Gamma_{PS}d_{c})(\bar{u}_d\overline\Gamma_{PS}\bar{d}^T_{e})\bigg)
\nonumber\\
&&+\frac{G}{4N_c(N_c+1)}(u^T_a\Gamma_{T,\rho\sigma}d_{a'})
\bigg((\bar{u}_a\overline{\Gamma}_T^{\rho\sigma}\bar{d}^T_{a'})
+(\bar{u}_{a'}\overline{\Gamma}_T^{\rho\sigma}\bar{d}^T_{a})\bigg)\ ,
\label{Ldi}
\end{eqnarray}
where the spin matrices are given by
$\Gamma_S=C\gamma^5$, $\Gamma_{PS}=C$, $\Gamma_{T,\rho\sigma}=C\sigma_{\rho\sigma}$
and $\overline{\Gamma}_i=\gamma^0\Gamma_i^\dagger\gamma^0$.
One can see that only three diquarks, scalar, pseudoscalar, and tensor diquarks
can strongly couple with the instanton. From the above Lagrangian,
for $N_C=3$, it is expected that the interpolating current with the coefficients,
$\alpha : \beta : \gamma=1:-1:1/4,\ \omega=\eta=0$,
may provide some specific properties to the OPE for the $\sigma(600)$.
Indeed, if we restrict our consideration only to the current of the scalar and the
pseudoscalar diquark--antidiquarks, we can immediately recognize that the equal weights
($\alpha^2=\beta^2$) between the two types of the diquarks give a special chirality
structure to the interpolating current :
\begin{eqnarray}
\alpha J_\sigma^{S\bar{S}}+\beta
J_\sigma^{P\bar{P}}&\sim&-(\alpha-\beta)(u_L^TCd_L\bar{u}_LC\bar{d}_L^T
+u_R^TCd_R\bar{u}_RC\bar{d}_R^T)
\nonumber\\
&&+(\alpha+\beta)(u_R^TCd_R\bar{u}_LC\bar{d}_L^T+u_L^TCd_L\bar{u}_RC\bar{d}_R^T)\ ,
\label{chirality}
\end{eqnarray}
where we have dropped the color indices
for simplicity. From this chirality structure, one can easily see that
the contributions from the operators associated with two chirality flips to
the OPE should have the numerical factor of $\alpha^2-\beta^2$. Therefore,
the dimension 8 operators, which gave the dominant contributions to the previous
sum rules with the scalar diquark--antidiquark only, will disappear in the sum rule
for the currents with the equal weights of the scalar and pseudoscalar
diquark--antidiquarks, $\alpha=\pm\beta$. The same cancelation happens in
the contributions from other high dimension operators and from the direct instantons
associated with the two chirality flips. Besides, the spin structure of the tensor
current restricts the OPE contributions from the high dimension operators.
As a result, one can expect stability in the sum rule
with such an interpolating current. By applying the two resonance
approximation to the phenomenological part of the sum rule to avoid
the well--known problem of the strong dependence of the multiquark mass
on the value of the threshold of the continuum~\cite{thres}, we get the mass of
$\sigma(600)$ around 780 MeV with the value of the threshold, $s_0=2.0$ GeV~\cite{Lee06}.
However, the numerical factor $\alpha^2-\beta^2$ appearing in the contributions from
the high dimension operators and from direct instantons may give indication on the existence
of two good interpolating currents for $\sigma$--state with $\alpha=\beta$ and
$\alpha=-\beta$. We have shown recently that the contribution arising from the two pion
intermediate state favors the tetraquark current with $\alpha=\beta$~\cite{Lee062}.

\section{Conclusions}
We have shown that the interpolating current consisting of the scalar
diquark--antidiquark and the pseudoscalar diquark--antidiquark for
the tetraquarks have very peculiar chirality structure. It has been demonstrated
that the chirality structure of the current with equal weights between the scalar
and the pseudoscalar diquark--antidiquarks leads to a vanishing dangerous contributions
from direct instantons as well as from the high dimension operators in the OPE.
As a result, we obtain the stable QCD sum rule and the mass of $\sigma(600)$
around 780 MeV.

\section*{Acknowledgements}
The author thanks the Yukawa Institute for Theoretical Physics at Kyoto University,
where this work was initiated during the YKIS2006 on "New Frontiers on QCD".
Also author thanks N.I. Kochelev for collaboration.



\begin{thebibliography}{99}
\bibitem{Amsler04}
C. Amsler and N.A. T\"{o}rnqvist, Phys. Rept, {\bf 384} (2004) 61.

\bibitem{smilga}
M.~S.~Dubovikov and A.~V.~Smilga, Nucl.\ Phys.\ {\bf B185} (1981) 109.

\bibitem{thres}
H.-J. Lee, N.I. Kochelev, and V. Vento,
Phys. Rev. D {\bf 73} (2006) 014010;
R.D. Matheus and S. Narison, arXiv:hep-ph/0412063.

\bibitem{jaffe2}
R.~L.~Jaffe and F.~Wilczek, Phys.\ Rev.\ Lett.\  {\bf 91} (2003) 232003.

\bibitem{shuryak}
E.~Shuryak and I.~Zahed, Phys.\ Lett.\  {\bf B589} (2004) 21.

\bibitem{Lee05}
H.-J. Lee, Eur. Phys. J. {\bf A30} (2006) 423.

\bibitem{Rapp98}
R. Rapp and T. Sch{\"{a}}fer, E. Shuryak, and M. Velkovsky, Phys.
Rev. Lett. {\bf 81} (1998) 53.

\bibitem{Lee06}
H.-J. Lee and N.I. Kochelev, Phys. Lett. {\bf B642} (2006) 358

\bibitem{Lee062}
H.-J. Lee and N.I. Kochelev, arXiv:hep-ph/0702225.
\end{thebibliography}
\end{document}